\documentclass[a4paper,10pt] {article}
\usepackage[latin1]{inputenc}
\usepackage{graphicx}
\begin{document}
\begin{titlepage}
\begin{center}
\Large{\bf Universal behavior in the static and dynamic properties
of the $\alpha$-XY model}
\end{center}
\vspace{1.5cm}
\begin{center}
\large{Andrea Giansanti$^{a, \S}$, Daniele Moroni$^{a, \dag}$ and
Alessandro Campa$^{b, \ddag}$}
\end{center}
\vspace{1cm}
\begin{center}
\normalsize{$^a$Dipartimento di Fisica, Universit\`a di Roma ``La Sapienza''
and INFM Unit\`a di Roma1, \\ Piazzale Aldo Moro 2, 00185 Roma, Italy}
\end{center}
\vspace{0.5cm}
\begin{center}
\normalsize{$^b$Laboratorio di Fisica, Istituto Superiore di Sanit\`a and
INFN Sezione di Roma1, Gruppo Collegato Sanit\`a\\Viale Regina
Elena 299, 00161 Roma, Italy}
\end{center}
\vspace{0.5cm}
\begin{center}
\emph{($8^{th}$ November 2000)}
\end{center}
\vspace{0.5cm}
\begin{center}
Accepted for publication in Chaos Solitons 
and Fractals:\\Special issue. Proceedings of the International Workshop:\\
{\it Classical and Quantum Complexity and Nonextensive Thermodynamics}, 
Denton (Texas), April 3-6 2000.
\end{center}
\vspace{1cm}
\begin{center}
\large{\bf Abstract}
\end{center}
\vspace{1cm}
\small{
The $\alpha$-XY model generalizes, through the introduction of a power-law
decaying potential, a well studied mean-field hamiltonian model with
attractive long-range interactions. In the $\alpha$-model, the interaction
between classical rotators on a lattice is gauged by the exponent $\alpha$
in the couplings decaying as $r^\alpha$, where $r$ are distances between
sites. We review and comment here a few recent results on the static and
dynamic properties of the $\alpha$-model. We discuss the appropriate
$\alpha$ dependent rescalings that map the canonical thermodynamics of the
$\alpha$-model into that of the mean field model. We also show that the
chaotic properties of the model, studied as a function of $\alpha$ display
a universal behaviour.
}

\vspace{0.5cm}

\noindent$^\S$\small{Corresponding author; Andrea.Giansanti@roma1.infn.it}\\
$^\ddag$\small{Campa@iss.infn.it}\\
$^\dag$\small{smoroni@lucifero.phys.uniroma1.it}\\
\end{titlepage}

\section{Introduction}
Equilibrium Classical Statistical Mechanics is one of the grounds of modern
physics; it originated from the works of Boltzmann and Gibbs at the end of
19th century. The computational techniques of Statistical Mechanics,
generalized to encompass the right enumeration of quantum states, have been
successfully applied to the study of all states of matter, along all the
20th century. On one side, the rigorous analysis of this branch of
theoretical physics has been impressive and far reaching \cite{ru}\cite{ga}
and the belief that everything in this field has been robustly founded
seems to be quite widespread. On the other side there has been a
serendipitous pragmatic use of the Boltzmann-Gibbs approach beyond the
bounds set by theorems.  The use of Equilibrium Statistical Mechanics
outside the allowed boundaries led to the discovery of anomalies or
paradoxes, particularly in the field of self-gravitating
systems\cite{lb}\cite{th}. In fact all the edifice of statistical mechanics
rests on few stringent assumptions on the interactions that are not
fullfilled by long-range forces. In spite of their fundamental relevance,
gravitational and Coulombic forces do not fit in with Equilibrium Classical
Statistical Mechanics\cite{note1}.

Short-range interactions guarantee extensivity: that is energy and entropy,
as functions of intensive internal parameters, grow linearly with the
number N of microscopic components of the systems. If an extensive system
is divided into macroscopic parts the total energy and entropy is the sum
of the energies and entropies of the parts. On the contrary that is not
true in systems where long-range interactions can reflect themselves in
thermodynamic potentials that do not scale with the size of the system and
cannot be defined in the thermodynamic limit (i. e.  for N going to
infinity). Moreover, long-range forces may well induce strong spatial and
temporal dynamic correlations that contrast mixing and result either in
very long relaxation times or in equilibrium distributions different from
the expected Boltzmann-Gibbs.

Aiming at a microscopic foundation of non extensive thermodynamics a good
starting point is the study of the ergodic properties in the thermodynamic
limit of simple and meaningful Hamiltonian models with long-range
interactions \cite{note2}. In the recent past Stefano Ruffo and co-workers
have introduced a class of mean-field models with infinite range of the
interactions \cite{art99}. These models are related with the statistical
physics of self-gravitating systems and that of phase transitions and
became paradigmatic in the study of non-extensivity. In particular, the
dynamics and thermodynamics of the so called Hamiltonian Mean-Field model
(HMF) have been extensively studied in the last few years and a
comprehensive review can be found in the contribution to these proceedings
by Andrea Rapisarda and Vito Latora \cite{LR}.

The $\alpha$-XY model we refer to in this paper has been introduced by
Celia Anteneodo and Constantino Tsallis \cite{AT}; this model cleverly
combines the physics of the mean-field models quoted above with the
statistical physics of lattice models of the Ising type with long-range
couplings decaying as the inverse power $\alpha$ of the distances between
sites. Since classical works on the foundations of Statistical Mechanics
\cite{fischer} it has been known that these systems are non extensive for
$0 \le\alpha/d < 1$, where $d$ is the dimensionality of the ambient space;
$\alpha/d$ is then the natural control parameter of non extensivity in this
class of models.

Our work originated from previous works by other researchers active in the
fields: in the section devoted to the thermodynamics of the $\alpha$-XY
model a canonical solution is presented that has been largely inspired by
the analytical work of Antoni and Ruffo on the HMF model \cite{AR} and by a
numerical work by Tamarit and Anteneodo \cite{TA} \cite{note3}; in the
section about dynamics the results on the Lyapunov exponents extend the
previous work by Anteneodo and Tsallis \cite{AT}. In the conclusions we try
to express our point of view about the connection of our results with the
particular form of nonextensive thermodynamics proposed by Tsallis
\cite{TsalProc}.

\section{Thermodynamics}
In this section we define the model and we compute its canonical partition
function. We use the typical methods of Gaussian transformation and saddle
point integration together with a Fourier diagonalization of the
interaction potential.

\subsection{Definition of the model}
We consider the hamiltonian of a system of bidimensional classical rotators
(XY spins):
\begin{equation} \label{num2.1}
 H=K+V=\frac{1}{2} \sum_{i=1}^{N} L_i^2 \, + \, \frac{1}{2\tilde{N}}
 \sum_{i\neq j}^{N} \frac {1-\cos(\theta_i-\theta_j)}
 {r_{ij}^\alpha} \, .
\end{equation}

The $N$ spins are placed at the sites of a generic $d$-dimensional lattice,
and each one is represented by the conjugate canonical pair
$(L_i,\theta_i)$, where the $L_i$'s are the angular momenta (unit momentum
of inertia is assumed) and the $\theta_i$'s $\in[0,2\pi)$ are the angles of
rotation on a family of parallel planes, each one defined at each lattice
point. The interaction between rotators $i$ and $j$ decays as the inverse
of their distance $r_{ij}$ to the power $\alpha\geq 0$. Periodic boundary
conditions are assumed in the form of nearest image convention.

The hamiltonian is extensive if the thermodynamic limit (TL)
$N\rightarrow\infty$ of the canonical partition function $(\ln Z)/N$ exists
and is finite. This is assured for each $\alpha$ by the presence of the
rescaling factor $\tilde{N}$ in front of the double sum of the potential
energy. $\tilde{N}$ is a function of the lattice parameters $\alpha,d,N$
which is proportional to the range $S$ of the interaction defined by:
\begin{equation}\label{range}
\tilde{N}\propto S=\sum_{j\neq i} \frac{1}{r_{ij}^\alpha} \, .
\end{equation}
The sum is independent of the origin $i$ because of periodic conditions.
Then, for each $\alpha$, $V$ is proportional to $N$. When $\alpha>d$, which
we call here the \textit{short-range} case, $S$ is finite in the TL
\cite{note2.1}, and things go as if each rotator interacted with a finite
number of rotators, those within range $S$. On the contrary when
$\alpha<d$, which we consequently call the \textit{long-range} case, $S$
diverges in the TL and the factor $1/\tilde{N}$ in (\ref{num2.1})
compensates for this.

The model (\ref{num2.1}) is intimately connected to its mean-field version
for $\alpha=0$, introduced by Antoni and Ruffo in \cite{AR} and called HMF
model:
\begin{equation} \label{num2.2}
 H=\frac{1}{2} \sum_{i=1}^N L_i^2 \, + \, \frac{1}{2N}
 \sum_{i,j=1}^N\left[1-\cos(\theta_i-\theta_j)\right]\, .
\end{equation}
The model is a classical ferromagnetic XY model and undergoes a
ferromagnetic transition from a high temperature paramagnetic state
crossing the critical temperature $T_c=1/2$ corresponding to a critical
energy $U_c=3/4$. The computation of the partition function of the full
$\alpha$ model, eq. (\ref{num2.1}) is a straightforward generalization of
the procedure followed for the HMF model, using the standard methods of
Gaussian transformation and saddle point technique, the whole treatment
being successful only for the long-range case $\alpha<d$. We resume here
the main passages, whereas a more detailed explanation can be found in
\cite{CGM}.

\subsection{Partition function}
The partition function of model (\ref{num2.1}) factorizes in a trivial
kinetic contribution $Z_K=(2\pi/\beta)^{N/2}$ and a nontrivial potential
part
\begin{equation}\label{num2.3}
Z_V=\int_{-\pi}^{\pi}d^N\theta\exp(-\beta V) \, ,
\end{equation}
where $\beta=1/(\kappa_B T)$ is the inverse temperature. We rewrite the
potential part as
\begin{equation}
V=\frac{1}{2\tilde{N}} \sum_{i\neq j}^{N} \frac {1-\cos(\theta_i-\theta_j)}
 {r_{ij}^\alpha} - \mathbf{h}\cdot\sum_{i=1}^N \mathbf{m}_i
\end{equation}
where we have introduced an external field ${\mathbf h}=(h_x,h_y)$ and
defined a local magnetization vector ${\mathbf m_i}=(m_{ix}, m_{iy})
=(\cos\theta_i,\sin\theta_i)$. The total magnetization is then given by
${\mathbf M}=(M_x,M_y)=\frac{1}{N}\sum_{i=1}^N\mathbf{m}_i$. The integral
(\ref{num2.3}) is remanipulated introducing matrix conventions. The
constraint $i\neq j$ over the double sum is removed defining $r_{ii}^\alpha
= 1/b$, a finite number. Since the numerator $1-\cos(\theta_i-\theta_j)$ is
zero for $i=j$ the choice of $b$ is free. The removal of the constraint
allows us to introduce the distance matrix $R_{ij}=\beta/(2
\tilde{N}r_{ij}^\alpha)$. After defining $\mathbf{B}=\beta \mathbf{h}$ and
$C=\exp[-\beta/(2 \tilde{N})\sum_{ij} (1/r_{ij}^\alpha)]$, the potential
part can be written as:
\begin{equation}\label{Zmatr}
Z_V=C \int_{-\pi}^{\pi}
 d^N\theta\, \exp\left[\sum_{i,j,\mu}m_{i\mu}R_{ij}
m_{j\mu} + \sum_iB_{\mu}m_{i\mu} \right],
\end{equation}
where $\mu=x,y$. Diagonalizing the distance matrix $R=(R_{ij})$ we can
write the first part of the exponent in (\ref{Zmatr}) in a suitable form
for application of gaussian transformation. The formula requires real and
positive eigenvalues: the first condition is satisfied because the matrix
is symmetric and the second by a proper choice of parameter $b$ because, as
it can be easily seen, the entire spectrum is linearly translated by it.

Application of the formula leads to known integrals in variables $\theta_i$
and the whole $Z_V$ can be rewritten as
\begin{eqnarray}\label{saddle}
Z_V&=&C\frac{\det R}{\pi^N}\int_{-\infty}^{+\infty} d^N\Psi_x
 d^N\Psi_y  \\
&& e^{ N\left[-\sum_{ij\mu} \Psi_{i\mu} \frac{R_{ij}}
 {N}\Psi_{j\mu} +\frac{1}{N}\sum_l \ln\left( 2\pi I_0\left( |2\sum_j R_{lj}
 \mathbf{\Psi}_j+\mathbf{B}|\right) \right) \right] } \nonumber \, ,
\end{eqnarray}
where $I_0$ is the zeroth-order modified Bessel function and the
$\mathbf{\Psi}_{i}$'s are the gaussian variables. Use of the saddle point
technique is now more delicate than in the mean-field case and requires a
careful study of the first and second derivative of the function in square
brackets at the exponential in eq. (\ref{saddle}), which we call $f(w)$,
where $w=(\Psi_{1x},\ldots,\Psi_{Nx},\Psi_{1y},\ldots,\Psi_{Ny})$. The
maximum is obtained for a vector $w_0=
(\Psi_{x},\ldots,\Psi_{x},\Psi_{y},\ldots,\Psi_{y})$, homogeneous on the
lattice sites. Defining ${\mathbf\Psi}= (\Psi_x,\Psi_y)$, its direction is
that of $\mathbf{B}$, and its modulus $\Psi$ is given by the solution of
the self-consistency equation:
\begin{equation}\label{psieq}
\Psi =\frac{I_1}{I_0}\left( \beta\left[A\Psi +h\right] \right) \, ,
\end{equation}
with
\begin{equation}\label{ascal}
A=\frac{1}{\tilde{N}} [ b + S]
=\frac{1}{\tilde{N}} \left[ b +
 \sum_{j\neq i}\frac{1}{r_{ij}^\alpha} \right] \, ,
\end{equation}
and where $I_1$ is the first order modified Bessel function. We note that
when $h=0$ we have infinitely many degenerate solutions, since only the
modulus $\Psi$ is determined.

Evaluation of the elements of the hessian matrix of $f(w)$ at the
stationary point just proves that it is a maximum. Full application of the
saddle-point method gives the potential energy contribution to the free
energy per particle $-\beta F$ in the TL as $f(w_0)$ plus a second-order
correction which can be estimated as
\begin{equation}\label{correction}
\lim_{N\rightarrow\infty}\frac{1}{N}
\sum_{{\mathbf k}}\ln[1-(\cdots)R_{\mathbf k}] \, ,
\end{equation}
where $R_{\mathbf k}$ are the eigenvalues of matrix $R$. Because of
translational invariance (in turn due to periodic conditions) they can be
computed through Fourier transform and thus labelled by the reciprocal
lattice vectors $\mathbf{k}$. The $\tilde{N}$ rescaling involved in the
definition of the matrix $R$ is the key feature to estimate the correction.
When $\alpha<d$ it is found that most of the eigenvalues go to zero in the
TL, the quantity being given by $N'$ such that $N'/N\rightarrow 0$. The
same estimate then applies to eq. (\ref{correction}) and the second-order
correction effectively vanishes. The procedure cannot be extendend to
$\alpha>d$ because no divergence occurs in the spectrum.

Once obtained the free energy, differentiation with respect to
thermodynamic variables gives the total magnetization $M$, which is found
to coincide with the gaussian variable $\mathbf{M}=\Psi$, and the internal
energy
\begin{equation}\label{energy}
U= \frac{1}{2\beta} + \frac{A}{2}(1-M^2) -hM \, .
\end{equation}

\subsection{Universality}
Equations (\ref{psieq}) and (\ref{energy}) have the exact form of the mean
field equations as soon as one takes $A=1$ or, otherwise stated, as soon as
the proportionality in eq. (\ref{range}) is replaced by an equality. As we
show below with this position in the long-range case the partition function
of the model then completely reduces to that of the mean-field case and
with it the equations of state, namely the magnetization curve $M\,vs\,U$,
and the caloric curve $T\,vs\,U$ where the internal energy is $U=<H/N>$.
The thermodynamic of the model thus becomes \textit{universal} in the sense
of being independent of the value of $\alpha,d$ for any $\alpha<d$. As a
consequence it shows a ferromagnetic transition at the universal critical
mean field energy $U_c=3/4$. There is a slight difference between an
equality in eq. (\ref{range}) and the position $A=1$, caused by the term
$b$ in eq. (\ref{ascal}) but it loses significance in the TL because of the
divergence of the second term in $S$. For finite $N$ it is a simple matter
of computation that the right definition to obtain universal state curves
is really
\begin{equation}\label{ntilde}
\tilde{N}=S=\sum_{j\neq i}\frac{1}{r_{ij}^\alpha} \, .
\end{equation}
Indeed physical equations cannot depend on a pure mathematical parameter,
such as $b$. Take for example the high energy form of the caloric equation
of state. Because the term $\cos(\theta_i-\theta_j)$ in eq. (\ref{num2.1})
is null on average (each rotator being approximately free), using
equipartition of energy we find $2U\sim T+S/\tilde{N}$. In order to
numerically obtain a universal equation of state we must make this
expression coincide with the HMF one $2U\sim T+1$ and equation
(\ref{ntilde}) is required.

\begin{figure}[ht]
\begin{center}
\includegraphics[bb=15 225 819 822,width=\textwidth,keepaspectratio]{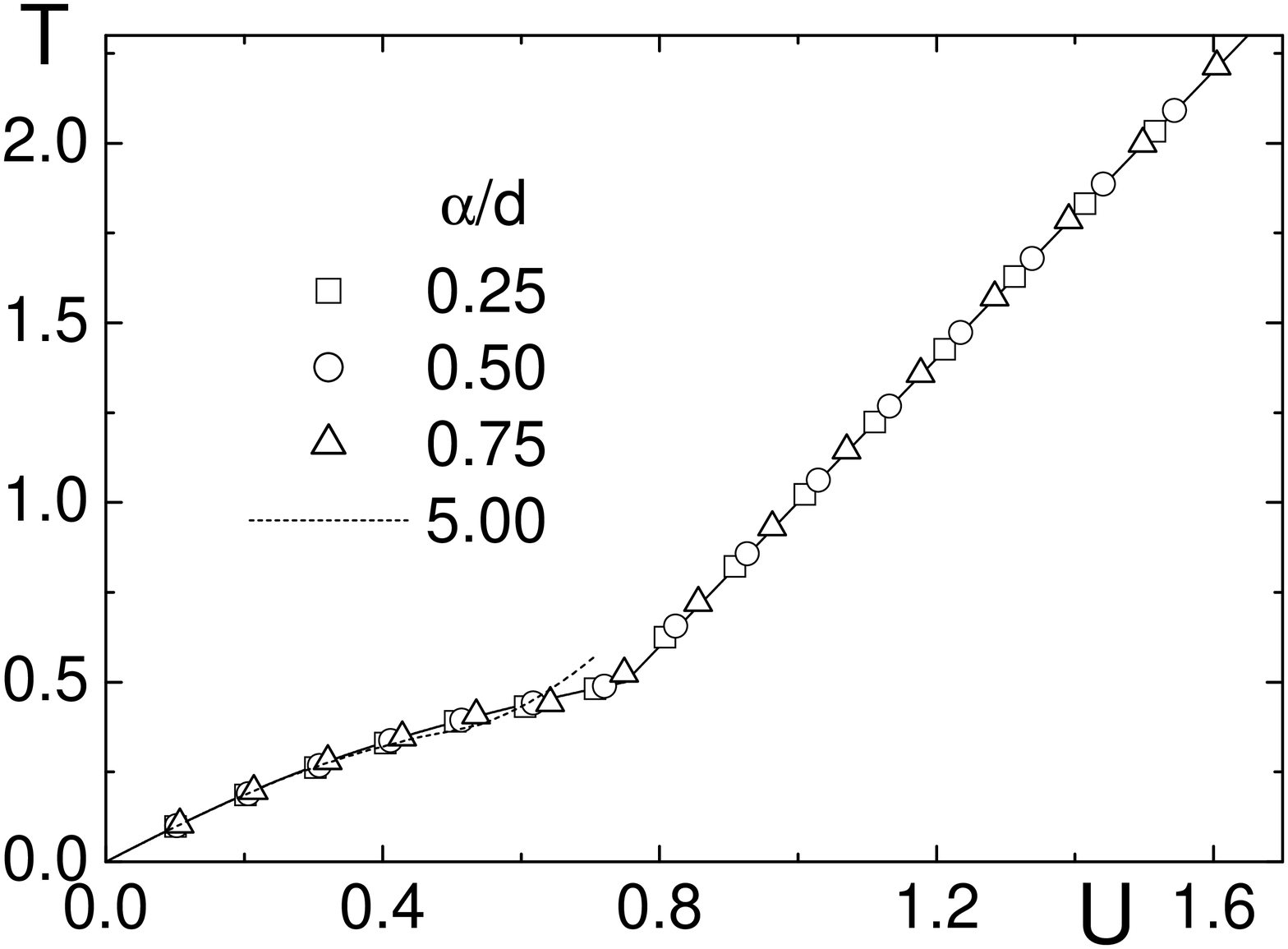}
\end{center}
\caption{The solid line gives the canonical theoretical caloric curve
(temperature $T$ $vs$ energy density $U$) for model (\ref{num2.1})
compared with microcanonical simulations of the model on a
threedimensional simple cubic lattice. Three values of $\alpha$ below $d$
(symbols) and one value above (dashed line) are shown.
Note the qualitative difference between the two cases.}
\label{fig.cal}
\end{figure}

\begin{figure}[ht]
\begin{center}
\includegraphics[bb=15 225 819 822,width=\textwidth,keepaspectratio]{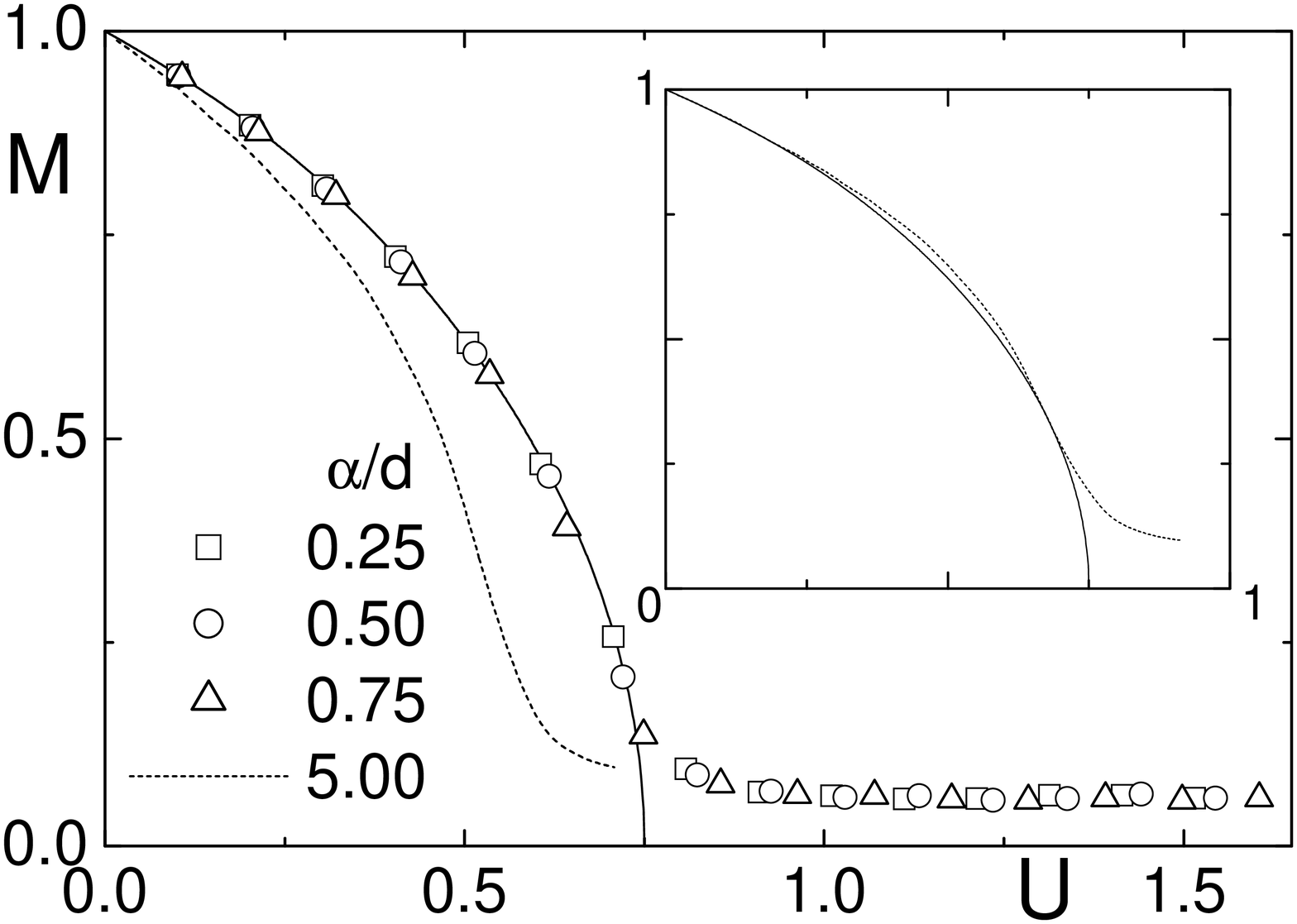}
\end{center}
\caption{
The solid line gives the canonical theoretical magnetization curve
(magnetization $M$ $vs$ energy density $U$) for
the system. The simulative points are represented by symbols for the
long-range ($\alpha<d$) cases and by the dashed line for the short-range
($\alpha>d$) case.
The difference between the long-range and the
short-range case is shown in the inset where the impossibility
for overlapping through abscissa rescaling is evident because of
a different critical exponent.}
\label{fig.magn}
\end{figure}

Then we have shown that any model with $\alpha<d$ on any lattice is
equivalent to HMF if we assume eq. (\ref{ntilde}). The theoretical
computation supports the simulative results found in \cite{TA} for a
onedimensional lattice. As a further check of the theoretical result we
simulated the system on a simple cubic three-dimensional lattice through
constant energy molecular dynamics applied to hamiltonian (\ref{num2.1}).
We used a fourth-order simplectic algorithm \cite{yosh} with time step
$0.02$, selected to have relative energy fluctuations not exceeding
$1/10^6$. We have chosen a fixed $N=343=7^3$, and have simulated various
energy densities $H/N$ and various $\alpha<3$. In Fig. \ref{fig.cal} we
show that the numerical caloric curves collapse onto the universal HMF
curve and in Fig. \ref{fig.magn} we show the same for the magnetization
curve. In the same figures we report simulations for one value of
$\alpha>d$ to stress the impossibility to extend the previous treatment to
the short-range case. The curves are qualitatively different: the caloric
one has an everywhere continuous derivative and lacks the finite
discontinuity at the critical energy, and the magnetization curve has a
different critical exponent. As the inset shows if one tries to rescale the
abscissa for the $\alpha>d$ magnetization curve in order to let it coincide
with the mean field one, a fundamentally different behaviour appears. As
known from renormalization group studies in fact the critical exponent for
the first-neighbour ($\alpha=\infty$) model is $\sim 0.36$ \cite{LT}
different from the mean-field value $0.5$.

It is also interesting to observe the extrapolation which follows from eq.
(\ref{ntilde}). If one estimates from this equation a \textit{mean} value
for $1/r_{ij}^\alpha$ one immediately finds
$1/r_{ij}^\alpha\sim\tilde{N}/N$. Insertion of this esteem into equation
(\ref{num2.1}) directly provides the fundamental HMF hamiltonian, eq.
(\ref{num2.2}). This remark naturally supports the previous analitycal
work, however we point out that it cannot fully justify it because it leads
to a wrong conclusion in the short-range case $\alpha>d$.

\begin{figure}[ht]
\begin{center}
\includegraphics[bb=15 15 819 582,width=\textwidth,keepaspectratio]{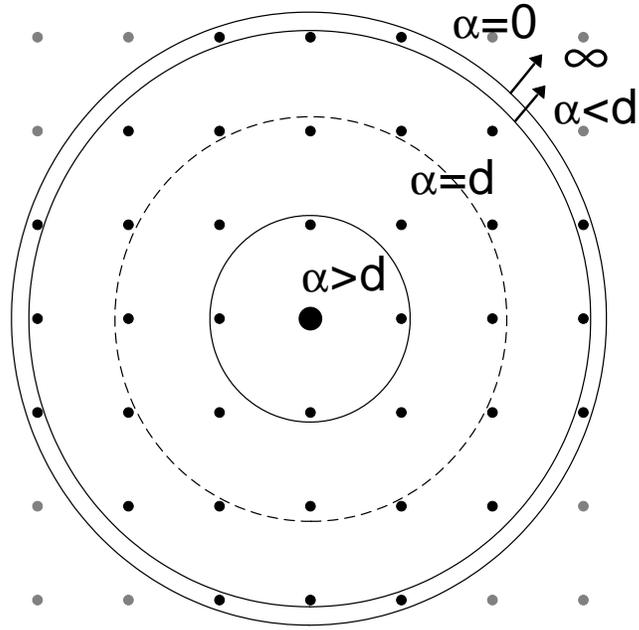}
\end{center}
\caption{\small Interaction area: it diverges for $\alpha<d$,
is finite for $\alpha>d$. The long range makes all areas
infinite and equal to each other.}
\label{goS}
\end{figure}

The physical reason for universality can be given considering a single
rotator surrounded by all others. It effectively interacts only with the
rotators within range $S$ (interaction area). For fixed $N$ and varying
$\alpha$ this area goes from all the lattice for $\alpha=0$ to the only
first neighbours for $\alpha=\infty$. Increasing $N$, if $\alpha<d$ the
range $S$ diverges and the interaction area expands as well, while if
$\alpha>d$ the range stays finite and the area localized. The idea is shown
in Fig. \ref{goS}. When $N\rightarrow\infty$ all interaction areas for
$\alpha<d$ simply become all the space, in the same way it occurs at finite
$N$ only for $\alpha=0$. The way the interaction areas diverge obviously
depends on $\alpha$ (see note \cite{note2.1}) but once at the infinity it
does not matter how you got there. All interaction areas are infinite and
equal to each other. Simply adapting the interaction field through
rescaling you obtain that each rotator feels all the others with the same
strength, independently of $\alpha,d$.

This reasoning cannot be extended to the case $\alpha>d$. Even for $N$
going to infinity the interaction area remains a well defined localized
zone, its features depending on $\alpha$ and the conformation of the
neighbourhood.

\section{Dynamics}

\subsection{Lyapunov exponents in the thermodynamic limit}
\begin{figure}[ht]
\begin{center}
\includegraphics[bb=15 15 819 582,width=\textwidth,keepaspectratio]{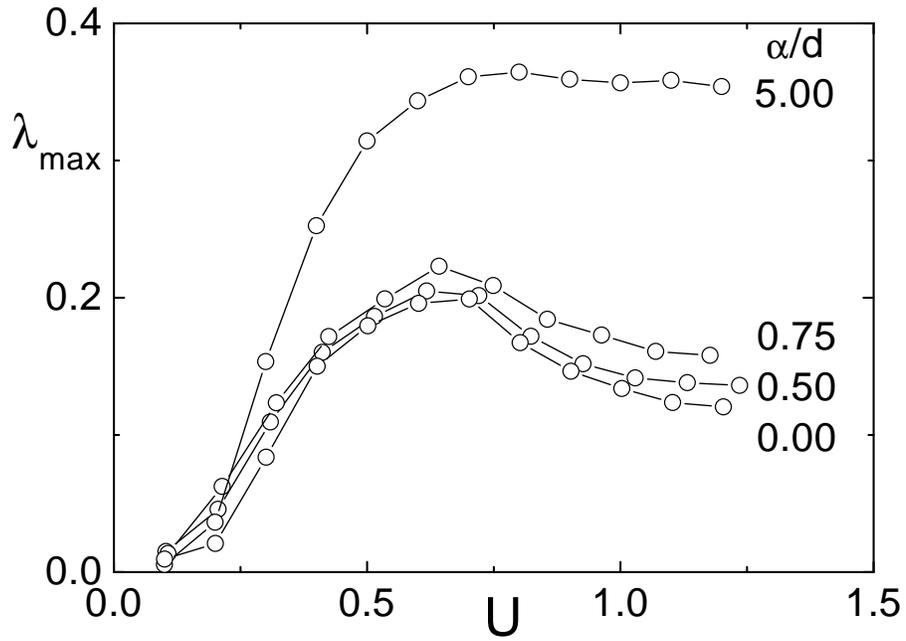}
\end{center}
\caption{\small Curves $\lambda_{max}(U)$ for $d=3$ ($N=343=7^3$) and
$\alpha/d=0,0.5,0.75,5$. Note the similarity between the long-range curves
for $\alpha<d$.}
\label{fig.lambdaU}
\end{figure}

We study the maximal Lyapunov exponent, as defined by the limit \cite{BGS}:
\begin{equation}\label{lambdat}
\lambda_{max}=\lim_{t\rightarrow\infty}\frac{1}{t}\ln
\frac{d(t)}{d(0)}=\lim_{t\rightarrow\infty}\lambda(t) \, ,
\end{equation}
with $d(t)=\sqrt{\sum_i (\delta \theta_i)^2 + (\delta L_i)^2}$ being the
metric distance calculated from infinitesimal displacements at time $t$
obtained in turn through the double integration of both the normal
equations of motion
\begin{eqnarray}
 \dot{\theta_i} &=& L_i \\ \label{eqmoto.a}
 \dot{L_i}&=&\sum_{j,j\neq i}
 \frac{\sin(\theta_j-\theta_i)}{\tilde{N}r_{ij}^\alpha} \label{eqmoto.b}
\end{eqnarray}
and the linearized ones
\begin{eqnarray}
\dot{\delta\theta_i}&=&\delta L_i \label{eqmototg.a} \\
\dot{\delta L_i}&=& \sum_{j\neq
i}\frac{\cos(\theta_j-\theta_i)}{\tilde{N}r_{ij}^\alpha}
(\delta\theta_j-\delta\theta_i)\label{eqmototg.b} \, ,
\end{eqnarray}
as obtained from hamiltonian (\ref{num2.1}). In the same simulations of the
previous section we also computed $\lambda_{max}$ constructing the curves
$\lambda_{max}(U)$ for various value of $\alpha$. We report the results in
Fig. \ref{fig.lambdaU}.

We note a common behaviour for all the curves at $\alpha<d$. Though not
perfectly coincident they all show a definite peak around the critical
energy $U_c=0.75$, resembling the HMF behaviour as found for example in
\cite{LRR1}\cite{LRR2}\cite{LRR3}\cite{LRR4}\cite{MCF}. On the contrary the
curve for $\alpha/d=5$ is different. Besides having a highest maximum, it
clearly lacks a well defined peak, and it much more resembles the
first-neighbour behaviour \cite{CCP}\cite{BC}.

We also investigated how $\lambda_{max}$ for varying $\alpha/d$ from $0$ to
$\infty$ goes from the HMF-like behaviour to the other one typical of the
first-neighbour model.

\begin{figure}[ht]
\begin{center}
\includegraphics[bb=15 15 819 582,width=\textwidth,keepaspectratio]{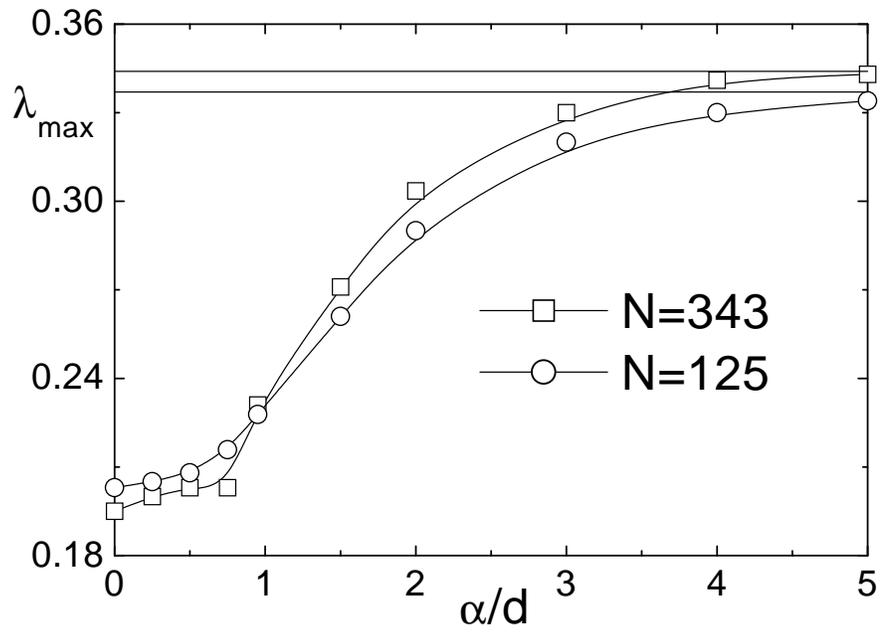}
\end{center}
\caption{\small $\lambda_{max}(\alpha/d)$ for $d=3$, $U=0.6$,
and $N=343=7^3,125=5^3$.
Solid curves are $B-spline$ interpolations to be used
as a guide for the eye. The long-range ($\alpha<d$) points are much closer than
the short-range ($\alpha>d$) ones. The horizontal lines are drawn at the
simulative values of $\lambda_{max}$ found
for $\alpha/d=\infty$ (first neighbour model).}
\label{lambdaU06}
\end{figure}

We simulated the system at fixed energy density $U_c=0.6$, varying
$\alpha/d$ and for $N=343$ and $N=125$. The results are shown in Fig.
\ref{lambdaU06}. In accordance with Fig. \ref{fig.lambdaU} the points for
$\alpha<d$ are much closer to each other, the vicinity increasing for
higher $N$. In the curve for $N=125$ a smooth ascent for low $\alpha/d$
suddenly increases after $\alpha/d=0.5$. The curve for $N=343$, though
showing a similar behaviour, presents the increase for $\alpha/d=0.75$. We
conjecture that for still higher $N$ the point of sharp increase moves
towards $\alpha/d=1$ and then in the TL we have
\begin{equation}\label{lambdauniv}
\lambda_{max}(\alpha,d)=\lambda_{max}(\alpha=0) \qquad \forall\:\alpha<d\,.
\end{equation}

\begin{figure}[ht]
\begin{center}
\includegraphics[bb=15 15 819 582,width=\textwidth,keepaspectratio]{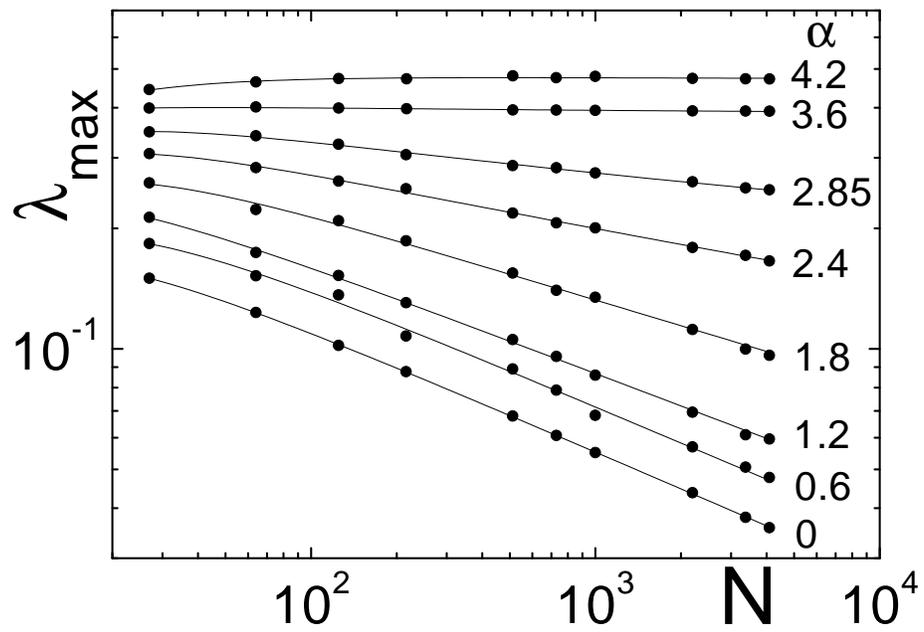}
\end{center}
\caption{\small $\lambda_{max}$ vs $N$ (log-log plot) for
$d=3$, $U=5$ and varying $\alpha$. Solid lines are fits with the
functional form $\left(a-\frac{b}{N}\right)/(\tilde{N})^c$
}
\label{liap3}
\end{figure}

We checked this conjecture with a more careful study for fixed high energy
$U=5.0>U_c$. The complete details can be found in \cite{CGMT}. These
simulations were done using a velocity-Verlet \cite{SABW} algorithm with a
time-step chosen to have a relative energy conservation of $10^{-4}$ or
better. Length of simulations were chosen looking at the asymptotic
behavior of quantity $\lambda(t)$, eq. (\ref{lambdat}), where $d(0)$ is
randomly chosen (see \cite{BGS}). We report in Fig. \ref{liap3}, the curves
$\lambda_{max}(N)$ for various $\alpha$ in dimensions $d=3$.

Because for the HMF model has been theoretically \cite{MCF} and
simulatively found that $\lambda_{max}(U>U_c,N\rightarrow\infty)=0$, it is
clearly seen that eq. (\ref{lambdauniv}) is satisfied. We observe that, for
growing $N$, if $\alpha>d$ then $\lambda_{max}$ is positive and constant,
whereas if $\alpha<d$ the maximal Lyapunov exponent tends to zero, the
value of HMF model. The same behaviour has been found in dimension $d=1$
\cite{AT} and $d=2$ \cite{CGMT}.

Then we can believe that in the TL the behaviour of the maximal Lyapunov
exponent is universal, its dependence on the energy density $U$ being the
same universal curve which can be found in Fig. 2 of \cite{MCF}.

\subsection{Finite $N$ case}

We have also investigated the scaling law for the reduction with $N$
of the Maximal Lyapunov Exponents. We report here the results 
shown in \cite{CGMT}.

For any  finite $N$ we have fitted the data with the following functional form:
${\lambda}_{max}\propto N^{-\kappa}$; $\kappa$ being the slope in the
log-log plots of Fig. \ref{liap3}. Collecting the slopes $\kappa$ as a
function of the ratio $\alpha/d$, for $d=1,2,3$, remarkably, through the
simple scaling $\alpha/d$, the $N$ dependence of the maximal Lyapunov
exponent of model (\ref{num2.1}) is universal. All curves start at
$\alpha=0$ from a value close to $1/3$, analytically and numerically found
for the HMF model \cite{LRR2}\cite{MCF}; when $\alpha/d$ increases from
zero to unity all our data collapse into a single curve, and then remain
zero for $\alpha/d$ greater than unity. And the universal function
$\kappa$, shown in Fig. \ref{kappa}, does not appear to depend on the
energy density, provided this is greater than the critical one.

\section{Quasi Stationary States}

In literature there is still another interesting feature of the HMF model
that deserves parallel study and generalization in the $\alpha-XY$ model.
Metastable states have been observed in microcanonical simulations that
give rise to anomalous thermodynamics and dynamics. In the range of energy
density $0.5<U<U_c=0.75$ states with an initial uniform distribution of
$L_i$'s (water bag distribution) have a slow relaxation and display a 
negative specific heat \cite{LRR2}\cite{LRR3}\cite{LRR5}\cite{LR00}. The average
kinetic energy, anomalously relaxes to a value which correspond to
a temperature value smaller than the canonical predicted one. The effect 
is purely microcanonical and
was also justified theoretically through the microcanonical solution of HMF
model \cite{AHR}. The velocity distribution of these states is not Gaussian
\cite{LR} and anomalous diffusion appears in the form of superdiffusion
\cite{LRR1}\cite{LRR4}\cite{LRR5}. These states are metastable and 
called \emph{quasi stationary states} because they have a lifetime which increases
with $N$. The lifetime of these states can be estimated either 
from the relaxation of temperature
to its canonical value or from the crossover of the mean square displacement 
from the superdiffusive to the
normal diffusive behaviour. The two estimates coincide and 
the lifetime is found to grow linearly with $N$ \cite{LRR5}.

When these states are considered the system then shows different
characteristics according to the order of limits in the computation of
observables. The idea is presented in Fig. 5 of \cite{TsalProc}. If the
thermodynamic limit is carried on before temporal average metastable
features sussist with their anomalies; on the contrary if the
$N\rightarrow\infty$ limit is taken as the second, normal canonical
prediction are valid and no superdiffusion is observed.

We believe these states are present also in the model we presented, eq.
(\ref{num2.1}). Accordingly we simulated the system at an energy density in
the above cited range and computed the velocity distribution function at
different times, see Fig. \ref{PDF}a, b. Though the numerical distribution
is constructed only through average on particles at fixed time and is not
therefore a smooth curve, it does not show a relaxation to the Gaussian
predicted form (continuous curve)
\begin{equation}\label{can}
F(L)=\frac{1}{\sqrt{2\pi T}}e^{-L^2/T}
\end{equation}
The logarithmic scale shows a difference in the tails of the distribution
resulting in a different microcanonical temperature (variance of the
simulative distribution) smaller than the canonical predicted one (variance
of the theoretical distribution). Probably the latter is reached for longer
integration times together with the gaussian form for the distribution. The
affirmation is currently under verification. We conjecture that Tsallis'
generalized thermostatistics may help to describe the features these states
show.

\section{Conclusions}
In the first part of this paper we have shown that the $\alpha$-XY model is
canonically equivalent to the HMF model. Through a rescaling by
$\tilde{N}$, as defined in (\ref{ntilde}), it is possible to map all
thermodynamic functions of the $\alpha$-XY model into that of HMF. All
$\alpha$-models have the same universal canonical thermodynamics, that of
the mean field model for $\alpha=0$. In these proceedings a microcanonical
solution of the HMF model has been given \cite{AHR}; a similar solution of
the $\alpha$-model will complete the proof of this kind of universality.

A dynamical equivalence between the HMF model and the $\alpha$-XY model is
also suggested by our initial exploration of quasi stationary states in
Fig. \ref{PDF}a, b. Let us note that what has been said for the
thermodynamic limit of these states in the HMF model should be extended to
the corresponding states of the $\alpha$-model.

\begin{figure}[ht]
\begin{center}
\includegraphics[bb=15 15 819 582,width=\textwidth,keepaspectratio]{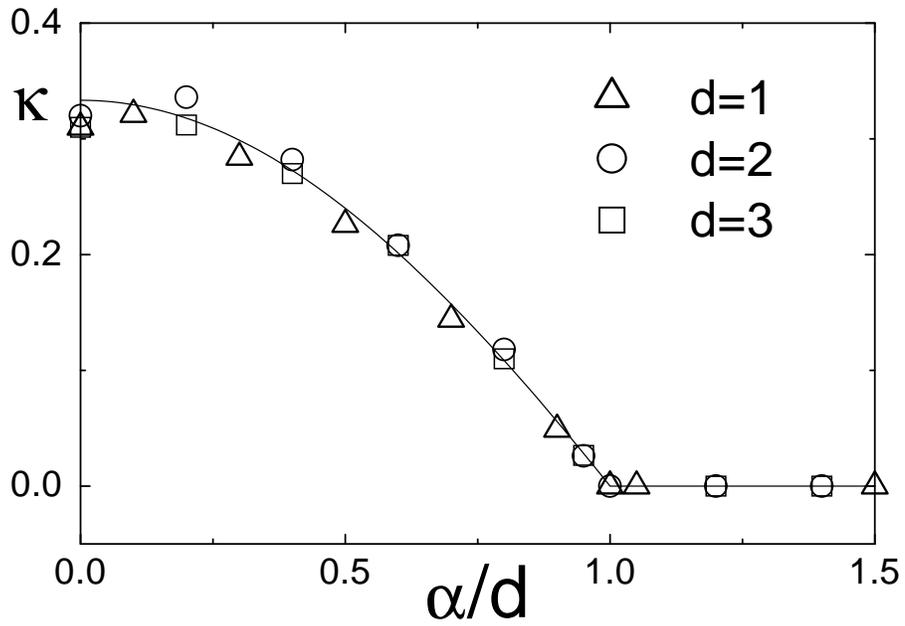}
\end{center}
\caption{\small The mixing weakening exponent $\kappa$ vs $\alpha/d$
for $d=1,2,3$ ($d=1$: from \protect\cite{AT}; $d=2,3$: from
\protect\cite{CGMT}).
It describes the asymptotic $N$ behaviour of the maximal Lyapunov exponent
$\lambda_{max}$ at fixed energy above the critical one, i.e.,
$\lambda_{max} \propto 1/N^{\kappa}$. The solid line is a guide to the eye
consistent with universality. For $\alpha=0$ we have
\protect\cite{MCF} $\kappa(0)=1/3 \; (\forall \, d)$.
Figure from \protect\cite{CGMT}.}
\label{kappa}
\end{figure}

\begin{figure}[ht]
\begin{center}
\includegraphics[bb=15 225 819 822,width=11cm,keepaspectratio]{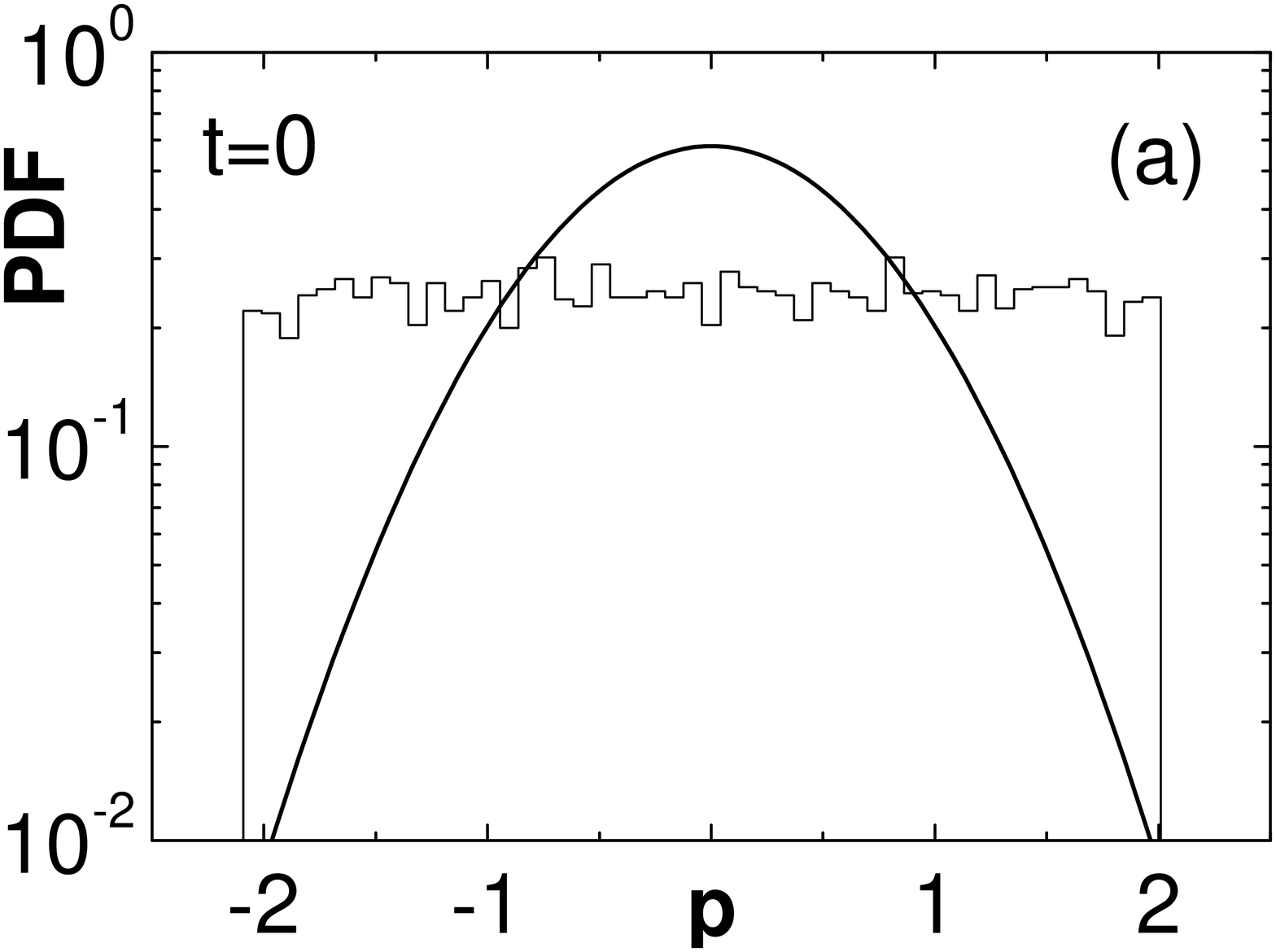}
\includegraphics[bb=15 225 819 822,width=11cm,keepaspectratio]{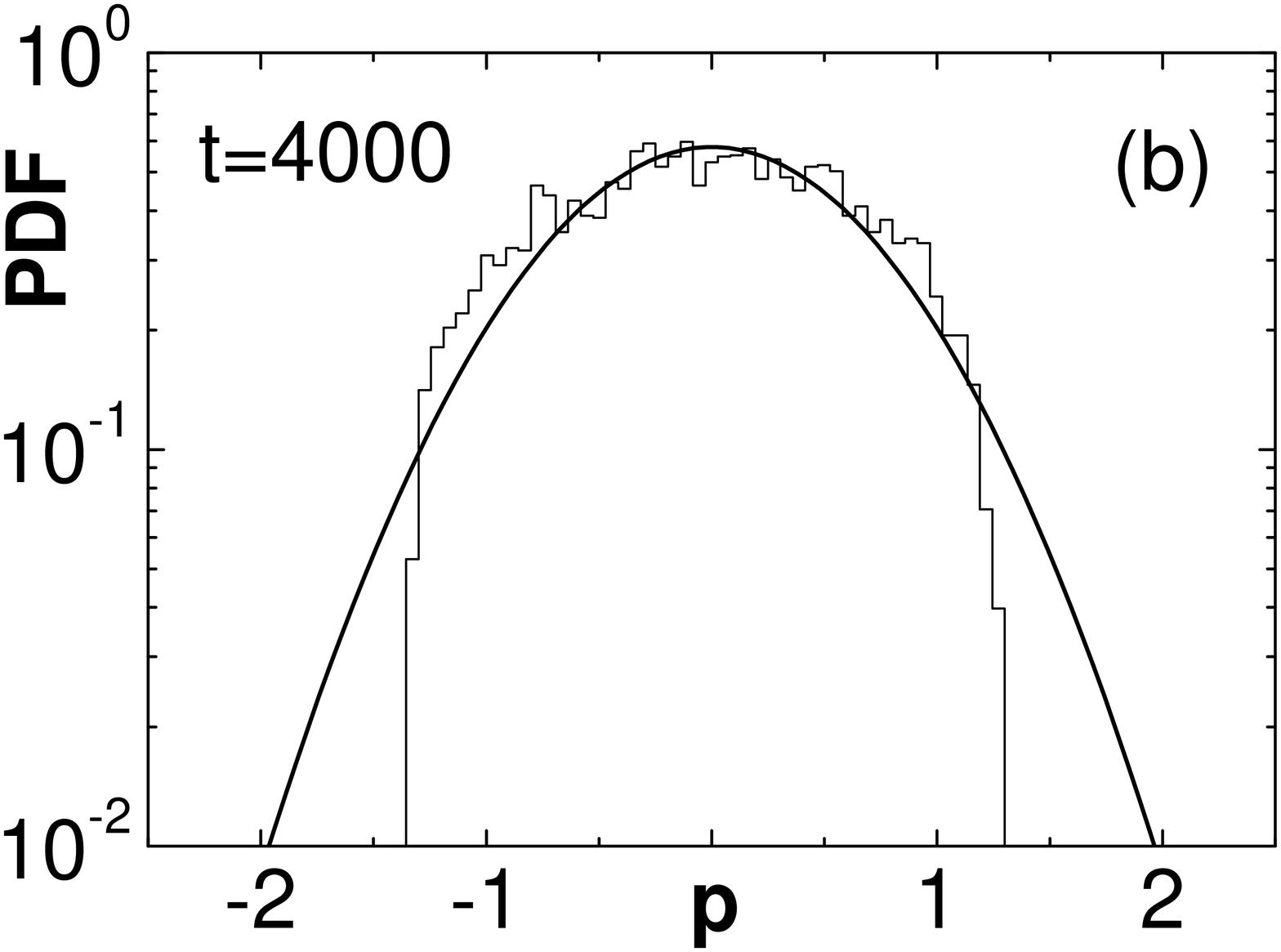}
\end{center}
\caption{Time evolution of probability distribution function (PDF) for
the momentum $p$ compared with
the theoretical canonical one. The graphs refer to a three dimensional
simple cubic lattice and $N=4096,\alpha=0.6,U=0.69$.
Water bag initial ($t=0$)
distribution is shown in (a). Microcanonical velocity distribution at $t=4000$
is shown in (b): the microcanonical temperature defined
as the variance of this distribution gives the value $T=0.39$.
continuous curves are the canonical distribution computed for $T=0.4757$
(see eq. (\ref{can})),
corresponding to $U=0.69$ in the canonical caloric curve,
see Fig. \ref{fig.cal}.}
\label{PDF}
\end{figure}

Another kind of universality has been found studying how maximal Lyapunov
exponents become zero in the thermodynamic limit in the nonextensive
$\alpha<d$ case. The curve $\kappa(\alpha/d)$ is universal; the way mixing
is reduced by the range of interaction in these $\alpha$-models is
controlled by the ratio $\alpha/d$. Firpo has given the theory of the
scaling of $\lambda_{max}$ with $N$ for the HMF model \cite{MCF}. It would
be nice to extend this theory to predict the universal form of the
$\kappa(\alpha/d)$ curve of our Fig. \ref{kappa}. If the reduction of
mixing here reported is connected with Tsallis' nonextensive thermodynamics
it should be possible to find out a strict relationship between Tsallis'
entropic index $q$ and $\alpha/d$; for $\alpha/d<1$ and searching at
equilibrium states in the limit $t\rightarrow\infty$ $N\rightarrow\infty$,
as discussed by Constantino Tsallis in his contribution.

Friendly and wit remarks by Stefano Ruffo have been very useful while we
were writing this review paper.



\begin{thebibliography}{2000}
\bibitem{note2.1}
The $N$ behaviour of the quantity $S$, eq. (\ref{range}) can be estimated
shifting to the integral $\int_1^{N^{1/d}} r^{d-1-\alpha} dr$
which gives $S\sim N^{1-\alpha/d}$.
\end{thebibliography}
\end{document}